\documentclass[aps,prd, twocolumn, nofootinbib, 10pt, preprintnumbers]{revtex4}
\usepackage{graphicx}
\usepackage{graphicx, epsfig}
\usepackage{color}
\usepackage{mathrsfs}
\usepackage{amsmath}
\usepackage{amssymb}
\usepackage{hyperref}

\newcommand{\beq}{\begin{equation}}
\newcommand{\eeq}{\end{equation}}
\newcommand{\bea}{\begin{eqnarray}}
\newcommand{\eea}{\end{eqnarray}}
\begin{document}
\title{The Overshoot Problem in Inflation after Tunneling}
\author{Koushik Dutta}
\author{Pascal M.~Vaudrevange}
\author{Alexander Westphal}
\affiliation{DESY, Notkestrasse 85, 22607 Hamburg, Germany}
\preprint{DESY 11-160}
\begin{abstract}
  We show the absence of the usual parametrically large overshoot
  problem of small-field inflation if initiated by a Coleman-De Luccia
  (CDL) tunneling transition from an earlier vacuum in the limit of
  small inflationary scale compared to the tunneling scale. For
  low-power monomial exit potentials $V(\phi)\sim\phi^n\;,\;n<4$, we
  derive an expression for the amount of overshoot. This is bounded
  from above by the width of the steep barrier traversed after
  emerging from tunneling and before reaching a slow-roll region of
  the potential. For $n\geq 4$ we show that overshooting is entirely
  absent. We extend this result through binomials to a general
  potential written as a series expansion, and to the case of
  arbitrary finite initial speed of the inflaton. This places the
  phase space of initial conditions for small-field and large-field
  inflation on the same footing in a landscape of string theory vacua
  populated via CDL tunneling.
\end{abstract}
\date{\today}
\maketitle
\section{Introduction}
The inflationary paradigm has become a part of the concordance model
of cosmology due to its huge number of observational post-dictions
\cite{Linde:2005ht}. Among other things, it can explain the flatness
of the universe, the lack of monopoles and the homogeneity, as well as
provide seeds of structure formation via perturbations of the inflaton
field, in excellent agreement with observations of the temperature
anisotropies in the cosmic microwave background (CMB)
\cite{Komatsu:2010fb}, \cite{Larson:2010gs}. An (as of now) unobserved
observational consequence of an inflationary epoch are primordial
gravitational waves (also known as tensor modes), imprinting
themselves in the B-mode polarization of the CMB. Unfortunately, their
amplitude is strongly model dependent. Inflationary models can be
broadly separated into two classes, large-field ($\Delta \phi > M_P$)
and small-field ($\Delta \phi < M_P$) models, giving rise to a small
and large amplitude of gravitational waves, respectively
\cite{Lyth:1996im}, where $\Delta \phi$ is the field variation during
inflation.

Despite its amazing success, the inflationary paradigm still faces
some important challenges. Many inflationary models suffer from a
sensitive dependence of the potential on the UV knowledge of our
theory. In addition, we still poorly understand the initial conditions
for inflation to start with. Successful inflationary models must
sustain a sufficient number efolds of inflation, usually taken to be
about $60$. For example in small-field models, if the scalar field
starts off at even a moderate distance uphill from the flat
inflationary plateau, or with an even very modest finite initial
downhill speed, it stays in the attractor solution for too little
time. In this case, the inflaton \emph{overshoots} the inflationary
plateau without ever settling into the slow-roll attractor
dynamics. This is the so-called overshoot problem first described
in~\cite{Brustein:1992nk}.

The problem of fine tuning the initial conditions has been studied
earlier. It has been found that for small-field models, e.g inflection
point inflation, the problem is rather acute \cite{Albrecht:1986pi},
\cite{Brandenberger:1988mc}, \cite{Goldwirth:1989pr},
\cite{Goldwirth:1990iq}. On the other hand, large-field inflationary
models are stable under moderate changes of initial conditions
\cite{Belinsky:1985zd}, \cite{Piran:1986dh}, \cite{Goldwirth:1989pr},
\cite{Brandenberger:1990wu}, \cite{Goldwirth:1990pm}.  For a review,
see \cite{Goldwirth:1991rj}. Recently, \cite{Bird:2008cp} discusses
the fine-tuning of initial conditions in terms of initial kinetic and
potential energy for a large class of single-field inflation
models. We will restrict our discussions of the overshoot problem to
small-field models, where the typical field range is sub-Planckian.

Various solutions to this problem have been proposed in the
literature, most often involving the effects of slower red-shifting
forms of matter-energy.  They start to dominate over the kinetic
energy of the scalar field and damp its motion by increasing the
Hubble friction. Examples include the use of additional fields
behaving like matter or radiation \cite{Kaloper:1991mq},
\cite{Brustein:2004jp} or primordial black holes coupling to the
scalar field~\cite{Kaloper:2004yj}. The role of non-zero momentum mode
in solving the overshoot problem has been pointed out in
\cite{Dine:2000ds}. It was also advocated that the string degrees of
freedom can generate a time dependent potential that can adibatically
track the field to the attractor \cite{Itzhaki:2007nk}. 

String theory is emerging as the prime candidate for a UV complete
theory of quantum gravity, making it compelling to search for
inflationary models in this context. At the same time, the standard
low energy approximation of the string theory leads to a `landscape'
of a large number of discrete vacua separated by potential barriers
\cite{Bousso:2000xa}, \cite{Susskind:2003kw}. Recently, there have
been a number of successful attempts at obtaining an inflationary
phase from constructions in string theory within this vast landscape
of true and false vacua. For some recent reviews see
e.g. \cite{Baumann:2009ni}, \cite{Cicoli:2011zz},
\cite{Burgess:2011fa}. 

The question of fine-tuning the initial conditions necessary to
achieve a sufficiently long period of slow-roll inflation in an
inflection point small-field potential from warped D3-brane inflation
in Type IIB string theory has been studied in \cite{Kachru:2003sx},
\cite{Baumann:2006th}, \cite{Krause:2007jk}, \cite{Baumann:2007ah},
\cite{Baumann:2008kq}. In this context, \cite{Underwood:2008dh} found
that almost the whole a priori microscopically allowed phase space of
initial conditions leads to successful inflation. Building upon
previous work by \cite{Baumann:2008kq}, it has been found in
\cite{Bird:2009pq} that the overshoot problem for warped D3-brane
inflection point inflation cannot be cured by DBI effects.

Finally,~\cite{Agarwal:2011wm} has further extended the analysis. They
allow for statistical distributions of the higher-dimension
corrections to the inflection point potential. Also, they widen the
initial condition phase space to include all 5 angular positions and
speeds of the D3-brane at the top of throat . This work found a
considerably reduced overshoot problem with respect to the
microscopically allowed initial conditions in the 12D phase space.
This may very well be due to the additional sources of matter-energy
stemming from including angular motion in the Hubble friction as well
as in the kinetic term for the radial position, i.e. the inflaton.

Instead of focusing on any particular string realization of inflation,
we will study the overshoot problem within the context of the
landscape of vacua in string theory. We would like to focus our
attention on the time immediately preceding the inflationary phase
that ended up being our local Hubble patch. At that time, the universe
can be assumed to be at a random position in the landscape, probably
in a false vacuum. After tunneling away from the false vacuum towards
the (hopefully true) vacuum, the inflaton field will generally appear
in a position away from the minimum. After rolling away from the
exiting part of the potential where the bubble tunnels, our Universe
must enter a (slow roll) phase that generates the observed $60$ or so
efolds of inflation, see e.g. \cite{Freivogel:2005vv}. A priori this
is not guaranteed, in particular, if the allowed field space in the
slow-roll plateau is sub-Planckian.

Within the field theoretic description, there are only two known
mechanism to traverse the landscape: CDL
tunneling~\cite{Coleman:1977py}, \cite{Coleman:1980aw} and
Hawking-Moss tunneling~\cite{Hawking:1981fz} for potentials which are
tuned to be very flat and wide. Within our setup, the CDL instanton is
dominant. Thus we assume that the evolution of the inflaton field
inside our Hubble patch was seeded by a CDL tunneling event with its
implicated boundary conditions. In particular, the CDL instanton
nucleates a bubble containing an infinite open universe with negative
spatial curvature and a scale factor $a(t)=t$ for $t\to 0$ where $t=0$
denotes the time of the tunneling event. At early times, the curvature
contribution to the Hubble friction is arbitrarily large. In other
words, the interior of the bubble is \emph{entirely} curvature
dominated at the very beginning, independent of the initial speed of
the scalar field(s).

It was realized first in~\cite{Freivogel:2005vv} for the simplified
example of a steep linear potential (mimicking the post-tunneling
steep downhill potential) matched to a constant piece (mimicking the
small-field inflation region) that this total curvature domination
ensures a finite overshoot independent of the potential at the exit
from tunneling (in the following we call this the `exit potential'). 
\begin{figure}
  \includegraphics[width=0.9\linewidth]{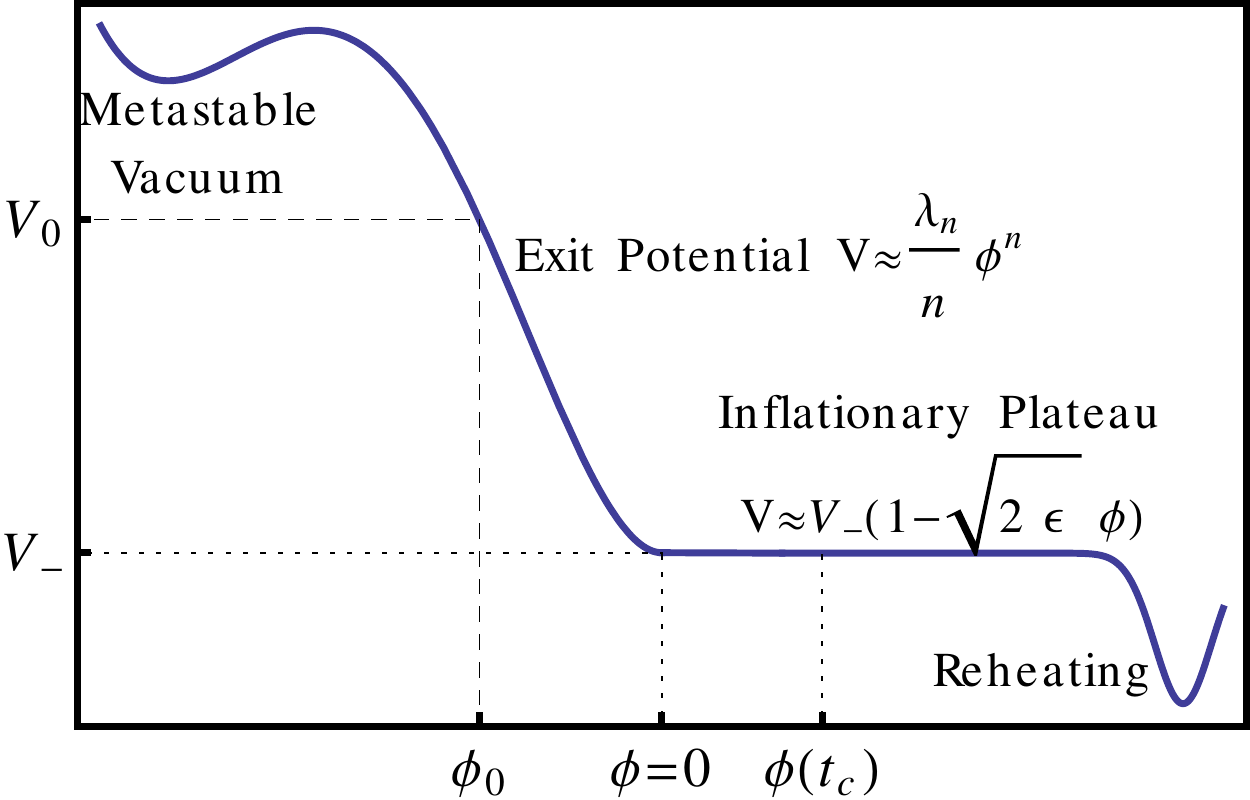}
  \caption{The overall set-up of our scenario. After tunneling from
    the metastable vacuum the inflaton field materializes at its exit
    point $\phi_0$ on the exit potential
    $V\sim\frac{\lambda_n}{n}\phi^n$. Then it rolls towards the
    minimum of the exit potential at $\phi=0$. Afterwards it enters
    the inflationary plateau with potential $V\sim
    V_-(1-\sqrt{2\epsilon}\phi)$. For $n=1,2,3$, the field will enter
    the inflationary phase at position $\phi(t_c)>0$. For $n\ge4$, the
    field will enter the inflationary phase already on the exit
    potential for $\phi<0$.}
  \label{fig:cartoon}
\end{figure}
Consider the sketch of the scalar potential given in
Figure~\ref{fig:cartoon}. Approximate the exit potential by a steep
linear downhill piece between $\phi_0<0$ and zero, and the
inflationary plateau by a constant piece for $\phi>0$. Due to
curvature domination at bubble nucleation, the scalar field can start
at arbitrarily high potential $V_0>> V_-$ (with $V_- $ the potential
of the constant inflationary part), and the field will reach slow-roll
speeds on the constant piece of the potential without parametrically
large overshoot. The field excursion $\Delta\phi_{CDL}$ between the
exit from tunneling and the onset of a potential dominated slow-roll
phase is limited and given in terms of of the distance $|\phi_0|$
between the exit from tunneling and the beginning of the plateau
region by~\cite{Freivogel:2005vv}
\begin{equation}
  \label{overshootLenny}
  \Delta\phi_{CDL}=2|\phi_0|\quad.
\end{equation}

Here we generalize this result to an arbitrary monomial potential
$V(\phi)=\frac{\lambda_n}{n}\phi^n$ describing the exit potential for
$\phi<0$. We chose to describe the inflationary plateau by a linear
potential for $\phi>0$ with slope $\epsilon$, the first slow-roll
parameter. We will describe this set-up in more detail in
section~\ref{Evolution_of_the_Universe_after_tunneling}. In
section~\ref{sec:monomials} we find that for the three lowest-power
monomial potentials $\phi,\phi^2,\phi^3$ describing the `exit
potential' from tunneling there is a finite amount of overshoot
\begin{equation}
  \label{overshootLena}
  \Delta\phi_{CDL}=\frac{3}{4\sqrt 2}\sqrt\epsilon+{\cal O}(1)|\phi_0|\quad,
\end{equation}
where the ${\cal O}(1)$ coefficient turns out to be a decreasing function of
$n$. 
\begin{figure}
  \includegraphics[width=\linewidth]{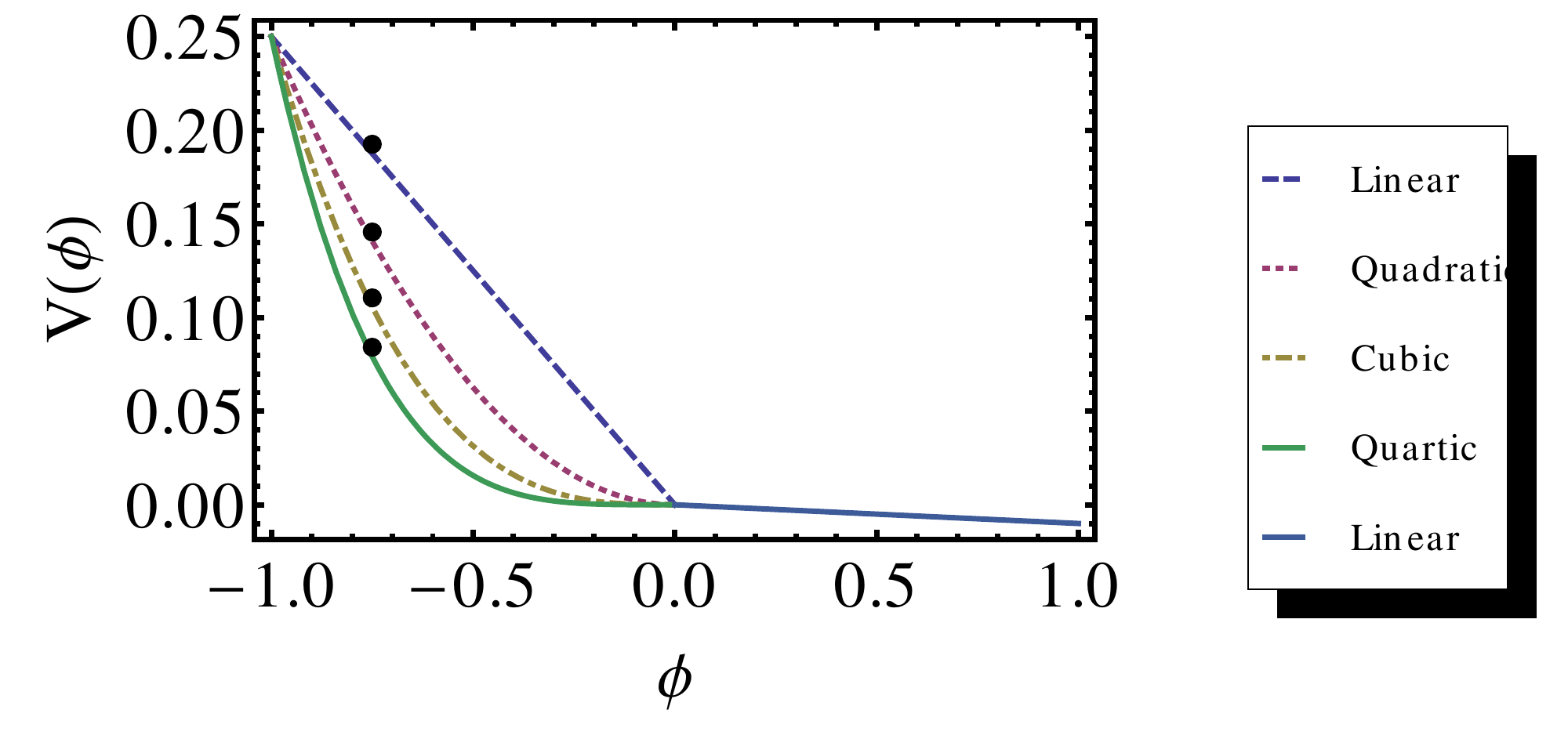}
  \caption{Plot of $V\sim\phi^n$ potentials. Increasing $n$ makes the
    potential steeper in the outer part to the left and flatter
    towards $\phi=0$.}
  \label{fig:Vplot}
\end{figure}
Surprisingly, for monomial potentials $V\sim\phi^n\;,\;n\geq 4$ there
is no overshoot at all. 

We can understand this intuitively in the following way. Increasing
$n$ increases the outer steepness of the potential, see
Figure~\ref{fig:Vplot}. This gets the field $\phi$ down into the
flatter part of the potential faster and therefore at larger Hubble
friction $\sim 1/t$. Thus the field spends more time with higher
friction in the flat part of the potential, leading to a more severe
slow-down. For large enough $n$, the field should come to a complete
stop. Thus we expect the overshoot for a given $\phi_0$ to decrease
with increasing $n$. Note that for expontential potentials,
\cite{Yamauchi:2011qq} showed the existence of tracking solutions
initiated in the early curvature dominated phase of the bubble. These
too track directly into slow-roll inflation on the inflationary
plateau.

In section~\ref{sec:polynomials} we generalize our analysis to a
binomial exit potential. It turns out that the overshoot depends on
$\phi_0$, like the monomials, and is mainly determined by the lowest
power monomial in the binomial. This property carries directly over to
a general exit potential written as a series expansion around the
point $\phi=0$ where slow-roll flatness has set in.

In section~\ref{sec:nonzero_initial_speed}, we show that this absence
of a parametrically large overshoot for small-field inflation is
actually \emph{independent} of the initial speed $\dot\phi_0$ of the
field. The only condition necessary for our results to hold are thus
that the CDL tunneling boundary conditions for the negative spatial
curvature and the scale factor are valid. However, this implies that
no-overshoot results generalize immediately to the case of CDL
tunneling in a multi-field landscape. There some of the scalar fields
will generically emerge with finite initial speed from the
instanton. 

In section~\ref{sec:conclusions} we conclude that the phase space of
initial conditions for small-field inflation and the one for
large-field inflation are on almost the same footing in a landscape of
string theory vacua populated via CDL tunneling.

\section{Evolution of the Universe after tunneling} \label{Evolution_of_the_Universe_after_tunneling}
In the String Theory ``landscape'' paradigm, it is conceivable that
our Universe originated via bubble nucleation from a nearby metastable
vacuum. This picture can be modeled by a single scalar field tunneling
from a false vacuum towards the true minimum. Before arriving at the
present minimum of our universe, whose potential energy is fixed by
the present value of the cosmological constant, the universe must pass
through a large vacuum energy dominated phase that can sustain about
60 e-folds of slow-roll inflation. If the vacuum energy is originating
from a scalar field potential, the field must have a shallow plateau
to sustain this inflation. Now, in the simplest case, the scalar field
which is responsible for creating our Universe through bubble
nucleation can also be identified as the inflaton. In this case the
bubble nucleating part of the scalar potential is followed by a
shallow plateau part. The potential energy of the plateau sets the
scale of inflation. The amount of inflation crucially depends on the
field range available on the plateau and on the initial speed of the
scalar field when it approaches the plateau. The initial speed depends
on the exiting part of the potential where the bubble nucleation
happens, as well as on the energy-matter content of the nucleated
bubble, in particular, contributions from the spatial curvature.

The cartoon in Fig.~\ref{fig:cartoon} describes the overall set-up of
our scenario where $\phi$ is the scalar field responsible for
tunneling and subsequent inflation. The scalar field tunnels from the
false vacuum towards $\phi_0$ with energy scale $V_0$ which for our
considerations is always much larger than the inflationary scale,
denoted by $V_-$. The tunneling part of the potential is attached to
the inflationary shallow potential at $\phi = 0$. The tunneling part
of the potential would be modeled by monomial shapes $V(\phi) =
\phi^n$ as well as polynomials. The plateau will be modeled by a
linear potential $V=V_-(1-\sqrt{2\epsilon}\phi)$. We will mainly focus
on analyzing the dynamics of the field $\phi$ for different powers of
the monomial.

For field distances larger than $M_P$, $i.e$ $\Delta\phi=|\phi_0| >
M_{P}$, between the nucleation point $\phi_0$ and the plateau starting
at $\phi=0$, the inflaton will be quickly slowed to its slow-roll
speed. Inflation will begin even before reaching to the plateau. In
this case, there is no over-shooting issue. The picture here is
similar to chaotic inflation or other large-field models for
super-Planckian field values.

On the other hand, if the allowed field range is sub-Planckian, $i.e$
$|\phi_0| < M_{P}$, as in small-field models, it is possible that the
field reaches the plateau with relatively high speed, potentially
overshooting the inflationary part. In this case, the field meets with
the inflationary attractor solution too late to sustain enough
inflation or does not reach to the attractor solution at all. One of
the main goals of this paper is to provide an estimate of this speed
for different forms of the exiting part of the potential and of the
amount of overshoot. Unless stated otherwise, we will assume
that the field range in this exiting part is always sub-Planckian. We
will find that the overshoot, while naively expected to be rather
severe, turns out be parametrically small or even zero.

Pioneered by Coleman and de Luccia, the process of bubble nucleation
is described by an instanton solution of the classical Euclidean
equations of motions. Inside the bubble, the universe is open with
negatively curved spatial hypersurface with curvature $1/a^2$, where
$a(t)$ is the scale factor of the Universe. We describe our Universe
by the standard FRW metric, and the equations of motion for the system
is given by
\begin{equation} \label{eq:eom:basic}
  \ddot \phi + 3 H \dot \phi = - V'(\phi)\quad,\nonumber\\
  H^2 = \frac{1}{3 M_P^2} \left(\frac{\dot \phi^2}{2} + V(\phi)\right ) - \frac{k}{a^2}\quad,
\end{equation}
where the Hubble constant $H = \dot a /a$ and $k = -1$. It is
important to note that when the bubble nucleates, $H$ is dominated by
the term $1/a^2$. Smoothness of the instanton requires the bubble to
start growing with initial conditions $a(t) = t + {\mathcal O}(t^3)$
and $\dot\phi(0)=0$. This in turn says that $ H = 1/t$ at the
very beginning, rendering the friction term in
Eq.~\eqref{eq:eom:basic} divergent. As a consequence, the field moves
very slowly in the beginning just after the field tunnels out. This
interesting feature due to the negative curvature of the spatial
metric inside the bubble has very important implications for the
overshoot problem that we are going to discuss soon.

\section{$\phi^n$ monomials with a linear slope attached}\label{sec:monomials}
In this section we will develop a detailed analysis of the dynamics of
the scalar field after it tunneled out of the metastable vacuum. For
simplicity we will assume now that the exiting part of the potential
can be approximated by monomials with different power $i.e$ $V(\phi)
\sim \frac{\lambda_n}{n}\phi^n$. The extremum of the monomial part,
chosen for convenience to be at $\phi = 0$, is attached to a shallow
linear slope on the right side (see Fig.~\ref{fig:cartoon}), and given
by
\begin{equation}\label{potential_right}
V(\phi)=V_-(1-\sqrt{2\epsilon}~\phi)\quad.
\end{equation}
This is the plateau of the potential and the scale is set by $V_-$. We
have written the potential in a form such that $\epsilon$ can be
easily identified with the inflationary slow-roll parameter given by
$\epsilon=(V'/ \sqrt{2}V)^2$.

We have already seen that the Universe is curvature dominated just
after the bubble materializes, thus $H = 1/t$. We will show later that
curvature dominates in all the monomial potentials until the field
reaches $\phi = 0$ as long as $|\phi_0| < M_P$, and $V_- << V(\phi_0)
= V_0$. Therefore we need to solve
\begin{eqnarray}\label{eq:eom:general}
  \ddot{\phi}+\frac{3}{t}\dot{\phi}+ (-1)^n \lambda_n \phi^{n-1}&=&0\quad,
\end{eqnarray}
subject to the boundary condition $\dot{\phi}(0)=0,
\phi(0)=\phi_0<0$. Note that any coefficient in the potential could be
absorbed in a rescaling of the time. The factor of $(-1)^n$ ensures
that the field always rolls from left to right.

We then use the solution of Eq.~\eqref{eq:eom:general} as a starting
point for solving the motion in the linear part on the right side
\begin{eqnarray}\label{eq:eom:linear}
 && \ddot{\phi}+3 H\dot{\phi}+\partial_\phi V=0\quad,
\end{eqnarray}
where $V(\phi)$ is given by Eq.~\eqref{potential_right}, and $V_-\ll
V(\phi_0)\equiv V_0$. At some time $t_c$ determined from $1/t_c^2 =
1/3V_-$, when the field is on the plateau, the curvature term becomes
subdominant compared to the potential. Once the potential starts to
dominate inflation sets in.

Before we perform a detailed analysis for the different forms of the
monomial potentials, we state the qualitative results that we are
going to establish in the remainder of this paper. There are two
qualitatively different classes of monomial potentials. For monomials
with power $n\ge4$, the field reaches the junction point $\phi = 0$
with zero velocity. In this case, potential energy even starts to
dominate before the field reaches the plateau. Therefore there is no
overshooting for $n\ge4$.  

On the other hand, for monomials with $n<4$, the field reaches the
junction point $\phi = 0$ with finite speed and will roll a certain,
parametrically small distance along the plateau the right until the
potential starts to dominate over the curvature at time $t_c$ and
inflation sets in.

It turns out that the distances $\phi(t_c)$ for the different values
of $n=1,2,3$ share some common features. Take the evolution of the
field on the linear slope on the right
\begin{eqnarray}
  \dot{\phi}+\frac{3}{t}\dot{\phi} - V_-\sqrt{2\epsilon}&=&0\quad,
\end{eqnarray}
and assume as initial conditions $\phi(t_f)=0,~
\dot{\phi}(t_f)=\dot{\phi}_f$, for some generic non-zero time $t_f$ and
initial speed $\dot{\phi}_f$. Then it is easy to see that at time $t_c$
$i.e.$ when inflation sets in, the field will be at a position
\begin{eqnarray}\label{general_final_position_1}
  \phi(t_c)&=&\frac{3}{4\sqrt{2}}\sqrt{\epsilon}+\frac{1}{2}\dot{\phi}_f t_f\\
  &&-\frac{1}{2\sqrt{2}}\sqrt{\epsilon}V_- t_f^2-\frac{1}{6}\dot{\phi}_f V_-t_f^3+\frac{1}{12\sqrt{2}}\sqrt{\epsilon}V_-^2 t_f^4\quad.\nonumber
\end{eqnarray}
Notice that the first term is independent of the parameters of the
potential on the left where tunneling happens which only impacts the
values of $\dot{\phi}_f$ and $t_f$.

In the subsequent analysis we will find that $t_f \sim
|\phi_0|/\sqrt{V_0}$, and $\dot{\phi}_f \sim \sqrt{V_0}$. Therefore
the second term in Eq.~\eqref{general_final_position_1} depends
parametrically only on the field value $\phi_0$ where the bubble
nucleated. Now in the approximation of $V_- << V_0$, it is easy to see
that all other terms are higher orders and therefore negligible. 

In summary, for a potential of form
$V(\phi)=\frac{\lambda_n}{n}\phi^n$ with $1 \leq n \leq 3$, the
overshoot is given by
\begin{eqnarray} \label{over_shoot}
  \phi(t_c) \simeq \frac{3}{4\sqrt{2}}\sqrt{\epsilon}+ \mathcal {O} (1) |\phi_0|\quad.
\end{eqnarray}
We will see later that the numerical prefactor in front of the
$\sqrt{\epsilon}$ is universal and independent of the power of the
monomials, whereas the coefficient of $\phi_0$ depends on the order of
the monomial, but is always of $\mathcal {O} (1)$. Now, the value of
$\epsilon$ is determined by the inflationary requirements, whereas the
value $\phi_0$ is calculable either exactly when the tunneling
potential is known \cite{Duncan:1992ai}, \cite{Hamazaki:1995dy},
\cite{Koyama:1999ai}, \cite{Dutta:2011ej}, \cite{Pastras:2011zr} or in
the thin-wall approximation \cite{Coleman:1977py}.

We note that the exiting part of the potential cannot be fully
approximated by monomials in a realistic setup of a false vacuum
separated from the true one by a potential barrier. The curvature of
the potential must change to make the turn around for the
barrier. But, as long as the the bubble nucleating point $\phi_0$ is
in a part where the curvature is positive, our analysis holds
fully. Now $V(\phi_0)$ must be smaller than the false vacuum height
for Coleman de Luccia instanton to be the consistent
solution~\cite{Coleman:1977py}, \cite{Coleman:1980aw}. Therefore,
unless the depth of the false vacuum is too shallow, the field will
tunnel out to a field value where the potential is positively curved,
making our approximation consistent. If the depth of the false vacuum
is small, the field will tunnel to a place where the potential is
negatively curved and some hill-top open inflation will ensue with
zero initial field velocity.

Also, in this case, the Hawking-Moss instanton might be the dominant
process instead of Coleman de Luccia
instanton~\cite{Hawking:1981fz}. We are not interested in this regime
and neglect it in the subsequent analysis. A realistic potential for
the tunneling part would consists of polynomials, instead of the
simplified assumption of a monomial that we are making now. In
section~\ref{sec:polynomials} we perform the analysis for polynomials,
finding results quantitatively very similar to overshooting dominated
by the lower monomial.

\subsection{$n=1$: The Linear}
Assuming curvature domination we can solve Eq.~\eqref{eq:eom:general}
for the linear potential
\begin{eqnarray}\label{eq:linear:eom}
  V(\phi)&=&V_-(1-\lambda_1\phi)\quad.
\end{eqnarray}
We find the monotonically growing solution in the exiting part of the
potential as
\begin{eqnarray}\label{eq:linear:phi_sol}
  \phi(t)&=&\phi_0+\frac{\lambda_1 V_-}{8}t^2\quad.
\end{eqnarray}
The field reaches at the bottom $\phi=0$ at
$t_f=2\sqrt{\frac{-2\phi_0}{\lambda_1 V_-}}$ with finite velocity
$\dot{\phi}(t_f) = \sqrt{-(\lambda_1 V_-\phi_0)/2}~.$

We need to justify the assumption of curvature domination by showing
that the contribution of curvature $3/t^2$ to the Friedman equation 
\begin{equation}
  H^2=\frac{1}{3}\left[\left(\frac{1}{2}\dot{\phi}^2+V(\phi)\right)+\frac{3}{t^2}\right]\quad,
\end{equation}
is larger than the other contributions. Plugging the solution
Eq~\eqref{eq:linear:phi_sol} in the expression for the total energy of
the scalar field, we obtain
\begin{eqnarray}
  \frac{1}{2}\dot{\phi}^2+V(\phi)&=&V_0-\frac{3}{32}t^2V_-^2\lambda_1^2\quad,
\end{eqnarray}
which is obviously smaller than $V_0$. As $t<t_f$ the curvature
contribution in this part of the potential $3/t^2$ is always larger
than $3/t_f^2$. Therefore to confirm the validity of our assumption we
just need to satisfy
\begin{eqnarray}
  V_0<\frac{3}{t_f^2}&\Rightarrow&1<\frac{3}{8\phi_0^2}\,\Rightarrow|\phi_0|<\frac{3}{8}\quad,
\end{eqnarray}
where we assumed that $V_-\ll V_0$. As it has been stated earlier, we
are solely discussing small-field models $|\phi_0| < M_P$, so that this condition on curvature domination is always fulfilled.

The finite velocity at the bottom lets the field roll up to 
\begin{eqnarray} \label{linear_overshoot}
  \phi(t_c)&=&\frac{3}{4\sqrt{2}}\sqrt{\epsilon}-\phi_0\quad,
\end{eqnarray}
before the vacuum energy $V_-$ of the plateau starts to dominate. The
subsequent field range on the plateau must be large enough to sustain
$60$ e-folds of inflation.  We remind the readers that this finding is
in agreement with our general result in Eq.~\eqref{over_shoot}. In the
limit of $\epsilon = 0$ where the plateau is fully horizontal, we
recover the known result of \cite{Freivogel:2005vv}: the field travels
a total distance from tunneling exit to beginning of inflation of $2
|\phi_0|$ in the field space.

\subsection{$n=2$: The Quadratic}
Now we discuss the situation when the exiting part of the potential
where the bubble nucleates is quadratic. Solving
Eq.~\eqref{eq:eom:general} for the quadratic potential
\begin{equation}
V(\phi) = V_- + \frac{1}{2} m^2 \phi^2\quad,
\end{equation} 
we find the solution for the field
\begin{eqnarray}
  \phi (t)&=&2\phi_0\frac{J_1(mt)}{mt}\quad, \label{sol:quadratic}
\end{eqnarray}
where $J_k(z)$ is the Bessel function of the first kind. The field
will reach $\phi=0$ at $t_f\approx3.83/m$ with a velocity of
\begin{eqnarray}
  \dot{\phi}(t_f)&=&\phi_0\frac{J_0(m t_f)-J_1(mt_f)}{t_f}\approx-0.21 m\phi_0\quad,
\end{eqnarray}
where we need to keep in mind that $\phi_0<0$. 

From the solution of Eq.~\eqref{sol:quadratic}, it is easy to show
that the potential and the kinetic energy contribution to the Hubble
equation is given by
\begin{equation} \label{quadratic_energy}
  \frac{1}{3}\left( V_- + \frac{2 \phi_0^2 (J_1(mt)^2 + J_2(mt)^2)}{t^2} \right)\quad,
\end{equation}
and for $V_{-} << V_0$, and $\phi_0 < M_P$, it is always subdominant
compared to the curvature contribution in the quadratic part of the
potential. This can be seen in the following way:

To neglect $V_-$ in Eq.~\eqref{quadratic_energy}, it is sufficient to
show the curvature term at the bottom of the quadratic being still
larger than $V_-$,
\begin{eqnarray}
  \frac{1}{3}V_-<\frac{3}{t_f^2}&=&\frac{3m^2}{3.83^2}\, \Rightarrow V_- < 0.61 m^2\quad.
\end{eqnarray}
Now we have assumed that $V_-\ll V_0$, or in other words
\begin{eqnarray}
  \frac{1}{2}m^2\phi_0^2&\gg& V_-\quad.
\end{eqnarray}
Now as $\phi_0^2<1$ in Planck units, $m^2\gg V_-$, satisfying the
above mentioned condition. Thus neglecting $V_-$, we note that
$|J_\nu(z)|\le\frac{1}{\sqrt{2}}$ for all $\nu\ge1, z\in \mathbf{R}$
(see 10.14.1 in \cite{DLMF}), making the total energy of the field
smaller than the curvature contribution of
$\frac{1}{t^2}.$\footnote{Alternatively, one could also plot the total
  energy of the scalar field as a function of the dimensionless
  quantity $z=mt$ to see the point.} Thus, assuming curvature
domination is consistent.

From the time $t_f$ on when the field reaches $\phi = 0$, it will roll
on the right linear slope until curvature becomes subdominant at
\begin{eqnarray}
  \phi(t_c)&=&\frac{3}{4\sqrt{2}}\sqrt{\epsilon}+\phi_0 J_0(m t_f)\\
  &=&0.53\sqrt{\epsilon}-0.403\phi_0\quad.
\end{eqnarray}

Let us compare this result with the corresponding result for the
linear case, Eq.~\eqref{linear_overshoot}. In the limit of $\epsilon =
0$, the overshooting is less for the quadratic case than the linear
slope. Also note that the prefactor of $\sqrt{\epsilon}$ turns out to
be the same for the linear and the quadratic potential, in accord with
our general result in Eq.~\eqref{over_shoot}.

\subsection{$n=3$: The Cubic}
Now we turn to the case when the monomial potential in the exiting
part is cubic. As far as we know, there is no known analytical
solution to the exact equation of motion
\begin{eqnarray}\label{eq:eom:cubic}
  \ddot{\phi}+\frac{3}{t}\dot{\phi}-\lambda_3 \phi^2&=&0\quad.
\end{eqnarray}
Therefore we will use the solutions of equations of motion with
slightly larger and smaller friction terms to obtain upper and lower
bounds on the terminal velocity.

We will prove now that the field reaches the bottom of the potential
$\phi = 0$ with non-zero speed. From Eq.~(4.24) in
\cite{ISI:A1985AQZ8100012}\footnote {Notice the different sign in
  front of $\phi^2$ which is compensated by flipping the sign of
  $\phi$ in the expression for the first integral Eq.~\eqref{eq:cubic:first_integral}.} we know that the equation with
slightly larger friction term than Eq.~\eqref{eq:eom:cubic}.
\begin{eqnarray}\label{eq:cubic:larger_friction}
  \ddot{\phi}+\frac{10}{3t}\dot{\phi}-\lambda_3\phi^2&=&0\quad,
\end{eqnarray}
has a first integral given by
\begin{equation}\label{eq:cubic:first_integral}
  C_0=\frac{1}{2}\dot{\phi}^2t^4+\frac{4}{3}\phi\dot{\phi}t^3-\frac{1}{3}\phi^3 t^4+\frac{2}{9}\phi^2 t^2-\frac{4}{9}\dot{\phi}t-\frac{28}{27}\phi\quad.
\end{equation}
Note that any coefficient in the potential can be
absorbed in a rescaling of the time, so without loss of generality we
set it to unity.

We can evaluate this first integral for the initial condition at $t=0,
\phi(0)=\phi_0,\dot{\phi}(0)=0$ to be 
\begin{eqnarray}
  C_0&=&-\frac{28}{27}\phi_0\quad.
\end{eqnarray}
When the field reaches the minimum at $\phi=0$, the first integral becomes
\begin{eqnarray}
  C_0&\equiv&-\frac{28}{27}\phi_0=\frac{1}{2}t^2\dot{\phi}^4-\frac{4}{9}\dot{\phi}t\quad.
\end{eqnarray}
leading to a non-zero terminal velocity $\dot{\phi_f}$ at $\phi=0$. 

As the friction term of this equation is larger than the friction term
of the equation we are actually trying to solve, the velocity at a
given position will be smaller than the velocity of the unknown
solution to Eq.~\eqref{eq:eom:cubic}. Thus even for the solution of
Eq.~\eqref{eq:eom:cubic}, the terminal velocity at $\phi=0$ is
non-zero.

\subsubsection{Lower Bound on the Terminal Velocity}
It is also possible to find an exact analytic solution to the equation
of motion of the field with a slightly larger friction term,
Eq.~\eqref{eq:cubic:larger_friction}. It is given by \cite{Polyanin}
\begin{eqnarray}
  \phi(t)&=&-\frac{2}{3t^2\lambda_3}\left\{z(t)^2\,\mathfrak{p}\left[\imath z(t),0,1\right]+1\right\}\quad,\nonumber
\end{eqnarray}
where $\mathfrak{p}(z, g_2, g_3)$ denotes the elliptic Weierstrass
function\footnote{To prove that this is a solution, we use the
  defining relation of the Weierstrass function:
  $\mathfrak{p}^{\prime}(z,g_2,g_3)^2=4\mathfrak{p}(z,g_2,g_3)^3-g_2\mathfrak{p}(z,g_2,g_3)-g_3$. Specifically,
  at its zero, we have
  $\mathfrak{p}'(\zeta)=\pm\sqrt{\frac{1}{81}\phi_0^2}$. To show that
  it fulfills the boundary condition, use l'H\^opital's rule.} and
$z(t)=\left(-\frac{3\lambda_3\xi t^2}{2}\right)^{1/6}$. Note we have reintroduced
$\lambda_3$ again.  This solution has $\dot\phi(0)=0$, and
$\phi(0)=\phi_0$ determines the integration constant
$\xi=28\phi_0$. The field will reach $\phi = 0$ when
\begin{eqnarray}
  \mathfrak{p}(\imath z,0,1)&=&-\frac{1}{z^2}\quad ,
\end{eqnarray}
$i.e$ when $z \approx 3.61$. This happens
at a time
\begin{equation}
  t_f \approx \frac{7.26}{\sqrt{-\lambda_3\phi_0}}\quad,
\end{equation}
with a velocity of 
\begin{eqnarray}\label{eq:cubic:larger_friction:final_velocity}
  \dot{\phi}(t_f)&\approx&-0.03~\phi_0\sqrt{-\lambda_3\phi_0}\quad .
\end{eqnarray}
Considering that we are working with a system that has larger friction
than the exact one, this estimation of terminal velocity is a lower
bound. We will comment at the end of the next subsection on the
validity of the assumption of curvature domination for the cubic
potential.

The finite velocity of
Eq.~\eqref{eq:cubic:larger_friction:final_velocity} at $\phi = 0$ will
lead to the field rolling up to
\begin{equation}
  \phi(t_c)=\frac{3}{4\sqrt{2}}\sqrt{\epsilon} - 0.1\phi_0\quad,
\end{equation}
again in accord with our expectations from
Eq.~\eqref{over_shoot}.

\subsubsection{Upper Bound on the Terminal Velocity}
Now we will estimate the upper bound on the velocity of the field when
it reaches at the minimum. It turns out that the case of a slightly
smaller friction term in the equation of motion
\begin{eqnarray}\label{eq:cubic:smaller_friction:eom}
  \ddot{\phi}+\frac{5}{3t}\dot{\phi}-\lambda_3\phi^2&=&0\quad,
\end{eqnarray}
also has a closed form solution~\cite{Polyanin}
\begin{eqnarray}
  \phi(t)&=&\frac{(8/3\lambda_3)^{1/3}}{t^{2/3}}\mathfrak{p}(t^{2/3}(3\lambda_3/8)^{1/3}+\zeta, 0,\xi)\quad,
\end{eqnarray}
where again $\mathfrak{p}(z, g_2, g_3)$ denotes the elliptic
Weierstrass function, and $\zeta,\xi$ are integration constants.

Imposing the initial conditions $\phi(0)=\phi_0<0$ and $\dot\phi(0)=0$ implies
\begin{eqnarray}
  \zeta&=&\int_0^\infty\frac{d x}{\sqrt{4x^3+\phi_0^2}}=2^{1/3}\frac{\Gamma(\frac{1}{3})\Gamma(\frac{7}{6})}{\sqrt{\pi}|\phi_0|^{1/3}}\quad,
\end{eqnarray}
and
\begin{equation}
\xi=-\phi_0^2\quad,
\end{equation}
where $\zeta$ denotes the first real zero of the Weierstrass function with
$\mathfrak{p}(\zeta,0,-\phi_0^2)=0$. 

The solution $\phi(t)$ reaches the extremum $\phi=0$ at the second zero
of $\mathfrak{p}(z,0,-\phi_0^2)$ at $z=2\zeta$, or
equivalently at $t_f=\sqrt{8/3\lambda_3}\zeta^{3/2}$. The
field velocity at that time is given by
\begin{eqnarray}\label{eq:cubic:smaller_friction:final_velocity}
  \dot{\phi}(t_f)&=&-\frac{\pi^{3/4}}{2\sqrt 3(\Gamma(\frac{1}{3})\Gamma(\frac{7}{6}))^{3/2}}\phi_0\sqrt{-\lambda_3\phi_0}\quad\nonumber\\
  &\approx&-0.174\phi_0\sqrt{-\lambda_3\phi_0}\quad.
\end{eqnarray}
Comparing this terminal velocity with the one for the smaller friction
term in Eq.~\eqref{eq:cubic:larger_friction:final_velocity}, we note
that they have the same $\phi_0$ and $\lambda_3$ dependence. This is
an upper bound for the terminal velocity of the cubic potential
Eq.~\eqref{eq:eom:cubic}.

Next we need to check for curvature domination throughout the phase
$\phi(t)<0$. For curvature to dominate, we need to have
\begin{eqnarray}\label{cubic_curvature_condition}
  \frac{1}{2}\dot{\phi}^2-\frac{\lambda_3}{3}\phi^3-\frac{3}{t^2}&<&0\quad.
\end{eqnarray}
Using the first integral for Eq.~(\ref{eq:cubic:smaller_friction:eom})
(see Eq.~(4.21) in \cite{ISI:A1985AQZ8100012}) while replacing
$t\rightarrow\frac{t}{\sqrt{\lambda_3}}$
\begin{eqnarray}
  C_1&=&\frac{1}{2}t^2\dot{\phi}^2+\frac{2}{3}t\phi\dot{\phi}-\frac{\lambda_3}{3}t^2\phi^3+\frac{2}{9}\phi^2\equiv\frac{2}{9}\phi_0^2\quad,
\end{eqnarray}
we can substitute the first two terms in
Eq.~\eqref{cubic_curvature_condition} to obtain the condition for
curvature domination as
\begin{eqnarray}
  \frac{2}{9}(\phi_0^2-\phi^2)-\frac{2}{3}\phi\dot{\phi}t-3&<&0\quad .
\end{eqnarray}
We are only dealing with $|\phi|<|\phi_0|<1$ in Planck units, so the
first term is always smaller than $\frac{2}{9}$. This gives us the
condition for curvature domination to be true if
\begin{eqnarray}\label{eq:cubic:smaller_friction:curvature_condition}
  -\frac{2}{3}\phi\dot{\phi} t&<&\frac{25}{9}\quad.
\end{eqnarray}
It is clear that both at the beginning (at $t=0$) and at the end (at
$\phi=0$) of the trajectory, this condition is fulfilled.

We now discuss the situation in between. We can estimate the maximum
velocity the field could have (admittedly a huge over-estimate) as the
velocity it would have at the bottom in case of zero Hubble
friction. The largest field value is $\phi_0$, and the time must be
smaller than $t_f$, giving
\begin{eqnarray}
  |\dot{\phi}|&<&\sqrt{\frac{2\lambda_3}{3}\phi^3} ,~~~~~~|\phi| < \phi_0\\
  t&<&t_f=\frac{4}{\sqrt 3}\frac{\left(\Gamma(1/3)\Gamma(7/6)\right)^{3/2}}{\pi^{3/4}\sqrt{-\lambda_3\phi_0}}\quad.
\end{eqnarray}
Plugging all the upper limits into
Eq.~(\ref{eq:cubic:smaller_friction:curvature_condition}), we obtain
\begin{eqnarray}
  \frac{8\sqrt 2}{9}\frac{\phi_0^2}{\pi^{3/4}}\left(\Gamma(1/3)\Gamma(7/6)\right)^{3/2}\approx2.09<\frac{25}{9}\quad.
\end{eqnarray}
We find the assumption of curvature domination to be correct for the
equation of motion with slightly smaller friction term. The velocities
for the solution of Eq.~\eqref{eq:eom:cubic} are even larger than the
velocities for a smaller friction term. Thus we proved that for
solutions of Eq.~\eqref{eq:eom:cubic}, curvature dominates at all
times, as long as $\phi_0<1$.

The final velocity of
Eq.~\eqref{eq:cubic:smaller_friction:final_velocity} lets the field
roll on the right side of the potential until
\begin{eqnarray}
  \phi(t_c) = \frac{3}{4\sqrt{2}}\sqrt{\epsilon}-\frac{1}{3}\phi_0\quad,
\end{eqnarray}
before the inflationary phase sets in, which is again in accord with
our findings in Eq.~(\ref{over_shoot}).

Thus we bounded the overshoot for the solution of the equation with
correct friction term $\frac{3}{t}$, Eq.~\eqref{eq:eom:cubic}. The
true value of the overshoot lies in between the one for the larger and
smaller friction terms, respectively
\begin{equation}
  \frac{3}{4\sqrt{2}}\sqrt{\epsilon}-\frac{1}{3}\phi_0 > \phi(t_c) > \frac{3}{4\sqrt{2}}\sqrt{\epsilon} - 0.1\phi_0\quad.
\end{equation}

\subsection{$n=4$: The Quartic}
We have seen that for monomial potentials up to order three, the field
always reaches the plateau at $\phi = 0$ with nonzero speed. For
successful inflation, the plateau needs to be sufficiently long as
there is always a chance of overshooting. In the last few subsections
we have calculated this amount of overshoot by estimating $\phi(t_c)$
where curvature becomes subdominant over the potential energy. 

We will show now that for the quartic potential, under the assumption
of curvature domination, the field arrives at the minimum with zero
speed, therefore providing ideal initial condition for inflation on
the plateau. In fact, curvature domination ceases even before the
field reaches the minimum at $\phi = 0$ and inflation takes place even
for $\phi < 0$.

For the quartic potential $V(\phi)=\frac{\lambda_4}{4}\phi^4$,
Eq.~(\ref{eq:eom:general}) with the boundary conditions
$\phi(0)=\phi_0, \dot{\phi}(0)=0$ is solved exactly by
\begin{eqnarray}\label{quarticsol}
  \phi(t)&=&\frac{8\phi_0}{8+t^2\lambda_4\phi_0^2}\quad .
\end{eqnarray}
This is a non-oscillating function that reaches the bottom of the
potential at $\phi = 0$ in infinite time $t_f\rightarrow\infty$.
Thus zero field velocity at the bottom of the potential will be zero
\begin{eqnarray}
  \dot{\phi}(t)&=&-\frac{16t\lambda\phi_0^3}{(8+t^2\lambda\phi_0^2)^2}\stackrel{t\rightarrow t_f}{\longrightarrow}0\quad.
\end{eqnarray}

We showed explicitly that there is no overshoot for the quartic
potential. The assumption of curvature domination will be justified in
section~\ref{On_curvature_domination}.

\subsection{Higher Order Monomials with $n > 4$}
To the best of our knowledge there is no closed solution of
Eq.~\eqref{eq:eom:general} for a monomial potential with larger than
quartic order subject to the boundary condition
$\dot{\phi}(0)=0$. However, we can make use of the fact that for a
differential equation of the form
\begin{eqnarray}\label{eq:eom:solvable}
  \ddot{\phi}+\frac{\gamma}{t}\dot{\phi}+ t^{\frac{\gamma(n-2)-(n+2)}{2}}\phi^{n-1}&=&0\quad,
\end{eqnarray}
there exists an expression for a first integral
\begin{eqnarray}\label{eq:first_integral}
  C_3&=&\frac{t^{\gamma-1}}{2}\left(\dot{\phi}^2t^2+\dot{\phi}\phi t(\gamma-1)\right)+\frac{\left(\phi t^\frac{\gamma-1}{2}\right)^{n}}{n}\quad,
\end{eqnarray}
with $C_3$ being a constant, see Eq.~(3.25) in
\cite{ISI:A1985AAR1000011}.  Note again that any coefficient
$\lambda_n$ in the potential $\sim \lambda_n\phi^n$ can be absorbed in
a rescaling of the time, so without loss of generality we set $\lambda_n=1$.

To be consistent with the general form of Eq.~\eqref{eq:eom:general},
the coefficient of $\phi^m$ in Eq.~\eqref{eq:eom:solvable} should be
constant, enforcing
\begin{eqnarray}\label{eq:condition:lambda_m}
  \gamma&=&\frac{n+2}{n-2}\quad.
\end{eqnarray}

As it turns out, Eq.~\eqref{eq:condition:lambda_m} holds exactly for
the quartic potential with $n=4$ and $\gamma=3$. Evaluating the
first integral of Eq.~\eqref{eq:first_integral} for the initial
conditions $\phi(0)=\phi_0, \dot{\phi}(0)=0$ at $t=0$ gives
\begin{eqnarray}
  C_3&=&0\quad.
\end{eqnarray}
In order to obtain the velocity of the field at the bottom of the
potential, we evaluate Eq.~\eqref{eq:first_integral} at the time $t>0$
when $\phi=0$, immediately giving
\begin{eqnarray}
  \dot{\phi}\Big|_{\phi=0}&=&0\quad,
\end{eqnarray}
in agreement with the results from the previous subsection.

In order to extract information about higher powers of the potential,
we observe that for Eq.~\eqref{eq:condition:lambda_m} to hold for
larger $n > 4$, $\gamma$ needs to go towards unity (for $n = 4\;\Rightarrow\;
\gamma = 3$, for $n>4\;\Rightarrow\;\gamma<3$). In this case, using the first
integral again, the terminal velocity at the bottom is also zero for
arbitrary $n\geq 4$. At the same time, we observe that the
friction term $\frac{\gamma}{t}\dot{\phi}$ in
Eq.~\eqref{eq:eom:solvable} for which the first integral holds is
always smaller than the true friction term $\frac{3}{t}\dot{\phi}$ for
powers $n>4$. Thus the rolling scalar field in the real model must
have a smaller field velocity than the velocity of the field in the
above toy model with smaller friction. As the field velocity at the
bottom with smaller friction is zero, field velocity with the true
friction must also be zero.

In summary, for a quartic or higher monomial potentials, there is no
overshoot. Assuming curvature domination, the field reaches
the bottom of the potential with zero velocity. In reality, the
potential energy of the inflationary plateau $V_-$ will take over
before the field reaches the minimum. Thus the field will at most have
slow roll speed when entering the linear part of the potential.

\subsection{On Curvature Domination for Higher Order Monomials} \label{On_curvature_domination}
So far, we have shown explicitly for $n = 1, 2, 3$ that the assumption
of curvature domination in the tunneling part of the potential is
fully consistent in our setup as long as $|\phi_0| < M_P$ and $V_- <<
V_0$. It remains to be shown for $n\ge4$ that curvature really
dominates until the field reaches the junction point at $\phi = 0$.
For potentials $\sim \phi^n$ with $n\ge4$, we can again make use of
the first integral $C_3$ in Eq.~\eqref{eq:first_integral}. We argue
that the true field motion must be slower than the one obtained from
Eq.~\eqref{eq:eom:solvable}, and in particular that the true $\phi(t)$
must be smaller than the solution of Eq.~\eqref{eq:eom:solvable}. Now
let us determine some properties of the latter solution. For this part
only, we deviate from our usual convention and assume that $\phi>0$,
i.e. the field is rolling from right to left.

Owing to the initial condition at $t=0, \dot{\phi}(0)=0$ and
$\phi(0)<\infty$, we know that $C_3=0$. We can solve
Eq.~\eqref{eq:first_integral} for $\dot{\phi}$
\begin{eqnarray}
  \dot{\phi}&=&-\frac{2}{n-2}\frac{\phi}{t}\pm\sqrt{\frac{4 \phi ^2}{(n-2)^2 t^2}-\frac{2 \phi ^n}{n}}\quad,
\end{eqnarray}
and notice that reality of $\dot{\phi}$ requires the discriminant to
be positive. Thus we have the requirement
\begin{eqnarray}
  \phi(t)&<&2^{1/(n-2)}\left(\frac{n}{(n-2)^2t^2}\right)^{1/(n-2)}\quad.
\end{eqnarray}
In particular, this means that the true solution must be even smaller
than this, and the potential $V\sim\phi^{n}$ will depend on time like
\begin{eqnarray}
  V(t)&<&2^{n/(n-2)}\left(\frac{n}{(n-2)^2t^2}\right)^{n/(n-2)}\quad,
\end{eqnarray}
which, for $n\ge4$, is always smaller than the curvature contribution
to the Friedman equation $\propto\frac{1}{t^2}$. In other words, if
curvature dominates over potential energy at some point (which is
definitely true as $t\rightarrow0$), then it will always dominate over
the potential energy.

\section{Analysis for polynomials} \label{sec:polynomials}

Polynomial potentials will generally not allow for an exact solution
of the corresponding equation of motion. We may, however, approximate
the full solution by gluing together the solutions of the monomials
dominating in a given interval of field values. We shall illustrate
this by the most simple yet non-trivial example of a binomial
\begin{equation}
V(\phi)=(-1)^m\frac{\lambda_m}{m}\phi^m+(-1)^n\frac{\lambda_n}{n}\phi^n\quad,\quad n>m\quad.
\end{equation}
We will see shortly that we need to consider only the situation where
$1\leq m\leq 4$.  As tunneling needs to inject the field at a
sub-Planckian displacement $|\phi_0|<M_P$ to avoid slow-roll directly
emerging from tunneling, we need to distinguish two cases.

In the first case, the higher power monomial has $n>4$ and thus we
have $\lambda_n=\tilde\lambda_n / \Lambda^{n-4}$ where $\Lambda\sim
M_P$ denotes the UV cutoff of the effective field theory used to
derive $V(\phi)$. Effective field theory then tells us -- barring
further information -- to assume $\tilde\lambda_n={\cal O}(1)$. This
implies a parametric suppression of the $n$-monomial for sub-Planckian
$\phi$ compared to the lower-order $m$-monomial, as soon as the
$n$-monomial corresponds an irrelevant operator ($n>4$). The dynamics
thus effectively reduces to the one described in the previous sections
for the lower-order $m$-monomial with $m\leq 4$.

The second case consists of the situation when both $m,n\leq 4$ and
thus the effective field theory argument from above cannot be used to
automatically suppress the higher-order term by sub-Planckian field
values. In this case, we may solve the equation of motion
approximately by gluing together the solutions $\phi_{L1}$ of the pure
$\phi^n$ and $\phi_{L2}$ of the pure $\phi^m$ potential, assuming
$n>m$. It is clear from
\begin{equation}
  \ddot\phi+\frac{3}{t}\,\dot\phi=-(-1)^m\lambda_m\phi^{m-1}-(-1)^n\lambda_n\phi^{n-1}\quad, n>m
\end{equation}
that the matching point $\phi_*$ may be approximated by the field
value where the accelerating 'force' $-V'(\phi)$ has equal
contributions from both monomials 
\begin{equation}
  \left.(-1)^m\lambda_m\phi^{m-1}\right|_{\phi=\phi_*}\cdot(1+\delta)=\left.(-1)^n\lambda_n\phi^{n-1}\right|_{\phi=\phi_*}\quad,
\end{equation}
where we allowed for perturbing this matching condition by
$1+\delta$. This will allow us to determine the quality of the
matching condition. Solving the matching equation yields
\begin{eqnarray}
  \phi_*&=&-\left[\lambda\,(1+\delta)\right]^{\frac{1}{n-m}}=\phi_*^{(0)}\cdot(1+\delta)^{\frac{1}{n-m}}\nonumber\\ &=&\phi_*^{(0)}\cdot\left(1+\frac{1}{n-m}\delta+{\cal  O}(\delta^2)\right)\quad,
\end{eqnarray} 
where $\lambda\equiv \lambda_m/\lambda_n$ and
$\phi_*^{(0)}=-\lambda^{\frac{1}{n-m}}$ is the matching field value
obtained for $\delta=0$. We can see that the matching field value is
perturbatively stable under moderate changes of the matching condition
itself.

As mentioned above, the solution $\phi_{L1,2}$ for $\phi<\phi_*,
\phi>\phi_*$ is given by the one for the pure monomial $\phi^n,
\phi^m$, matched at
\begin{equation}
  \phi_{L1} = \phi_{L2}=\phi_*, \quad \phi_{L2}(t_*) = \phi_{L1}(t_*)\quad,
\end{equation}
where $t_*$ is defined as $t_*\equiv t(\phi_{L1}=\phi_*)$. $\phi_{L1}$
will be given by the solution to either a quadratic, cubic, or quartic
monomial with $\phi_{L2}$ the solution of the monomial of lower order,
i.e. cubic, quadratic or linear. 

The case $n=2$ is special, as it leaves only $m=1$. This can be
treated in the ensuing equation of motion exactly be shifting
$\phi\to\phi-\lambda_2/\lambda_1$.

We will now use the case $n=4$ and $m =1$ with its convenient
solutions of Eq.~\eqref{quarticsol} and Eq.~\eqref{eq:linear:phi_sol}
respectively as an example to demonstrate the parametric sensibility
of $t_*$ and $\dot\phi_{L1}(t_*)$ on perturbing the matching condition
by $\delta$. Demanding from matching that $\phi_{L1}(t_*)=\phi_*$ we
get
\begin{eqnarray}\label{matchings}
  t_*&=&\sqrt{\frac{8}{\lambda_4\phi_0^2}\,\left(\frac{\phi_0}{\phi_*^{(0)}}\,\frac{1}{(1+\delta)^{1/3}}-1\right)}\nonumber\\
  &=&t_*^{(0)}\cdot\left(1+\frac{\phi_0}{6\,\big(\phi_*^{(0)}-\phi_0\big)}\,\delta+{\cal O}(\delta^2)\right)\,,\nonumber\\
  \dot\phi_{L1}(t_*)&=&\frac{\sqrt{\lambda_4}}{4}\big(\phi_*^{(0)})^2(1+\delta)^{2/3}\\
  &&\times \sqrt{8\,\left(\frac{\phi_0}{\phi_*^{(0)}}\,\frac{1}{(1+\delta)^{1/3}}-1\right)}\nonumber\\
  &=&\dot\phi_L\big(t_*^{(0)}\big)\cdot\left(1+\frac{\phi_0-\frac43\phi_*^{(0)}}{2\big(\phi_0-\phi_*^{(0)}\big)}\,\delta+{\cal O}(\delta^2)\right)\quad.\nonumber
\end{eqnarray}
We see that the initial conditions for the subsequent
$\phi_{L2}$-evolution towards $\phi=0$, $t_*$ and $\dot\phi_L(t_*)$,
are stable under moderate perturbations of the matching condition. 

We expect qualitatively similar behavior for other values of $m, n$.
The solutions for all values of $m,n$ are polynomial,
i.e. non-exponential, in all parameters. This should translate into
perturbative stability of the final speed $\dot\phi_f(\phi=0)$ under
moderate changes of the matching condition.

\subsection{Binomial $V=V_{-}-\lambda_1\phi+\frac{1}{4}\lambda_4\phi^4$}
\begin{figure*}
  (a)\includegraphics[width=0.45\textwidth]{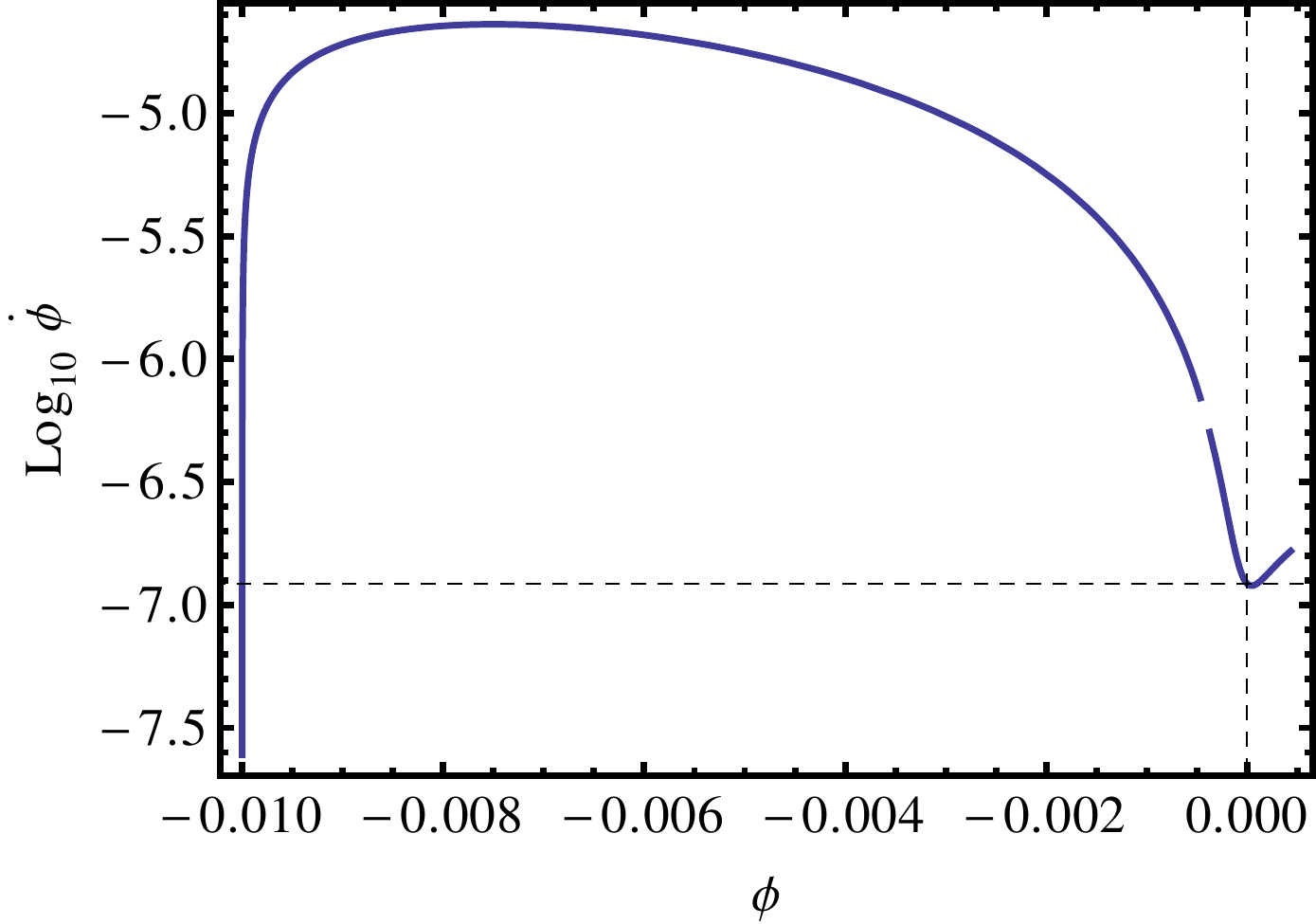}
  (b)\includegraphics[width=0.45\textwidth]{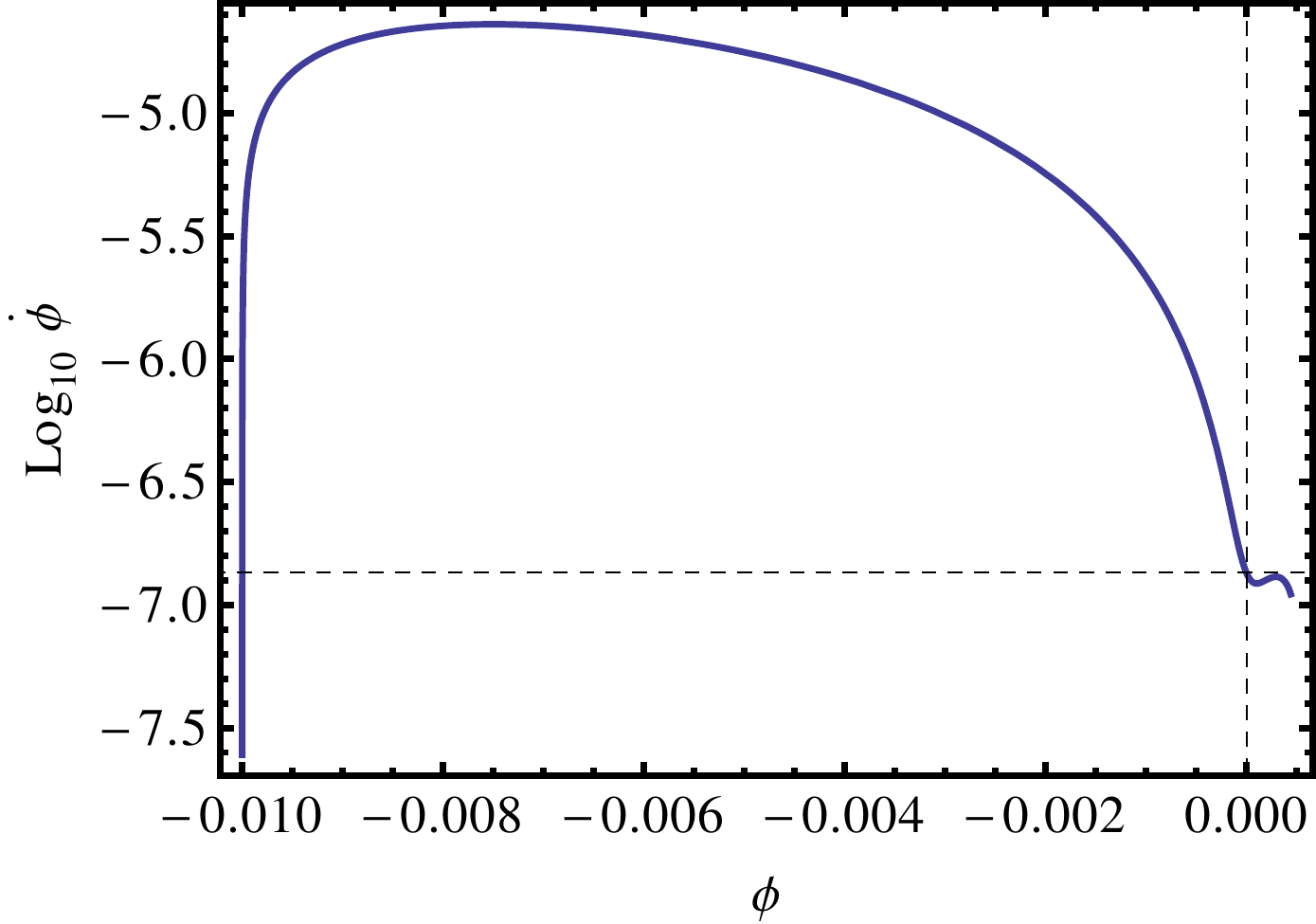}
  \caption{Phase space for a field rolling in the potential
    $V=-\lambda_1\phi+\frac{1}{4}\lambda_2\phi^4$ (a) using the
    analytical expression Eq.~\eqref{eq:linear_plus_quartic:analytic} and
    (b) using numerical integration of the differential equation. The
    analytical approximation agrees quite well with the results from
    numerics. The terminal speed is not clearly visible in this plot,
    but evaluates to $\dot{\phi}(t_f)=1.22\times10^{-7}$ (analytic)
    and $\dot{\phi}(t_f)=1.35\times10^{-7}$ (numerical).}
  \label{fig:linear_plus_quartic:phase_space}
\end{figure*}

We will now proceed to analytically estimate the terminal velocity at
$\phi = 0$ for the potential
$V=V_{-}-\lambda_1\phi+\frac{1}{4}\lambda_4\phi^4$.  As outlined
above, we assume that at early times $t<t_*$, $\lambda_4|\phi|^3 >
\lambda_1\,,$ $i.e$ the field evolution is dominated by the quartic
term, while for $t > t_*$ it is determined by the linear term.  The
following equations of motions
\begin{eqnarray}
  \ddot{\phi}_{L1}+\frac{3}{t}\dot{\phi}_{L1}+\lambda_4 \phi_{L1}^3&=&0\quad,\\
  \ddot{\phi}_{L2}+\frac{3}{t}\dot{\phi}_{L2} - \lambda_1&=&0\quad,
\end{eqnarray}
are subject to the boundary conditions
\begin{eqnarray}
  \dot{\phi}_{L1}(0)=0\,&,\,&\phi_{L1}(0)=\phi_0\quad,\\
  \dot{\phi}_{L2}(t_*)=\dot{\phi}_{L1}(t_*)&,\,&\phi_{L2}(t_*)=\phi_L(t_*)\quad,
\end{eqnarray}
are solved by
\begin{eqnarray}\label{eq:linear_plus_quartic:analytic}
  \phi_{L1}(t)&=&\frac{8\phi_0}{8+\lambda_4\phi_0^2\, t^2}\quad,\\
  \phi_{L2}(t)&=&\frac{\lambda\,\lambda_4}{8}t^2+\frac{8(\lambda^{1/3}+\phi_0)^3}{\lambda_4\phi_0^4}t^{-2}+\frac{2\lambda+3\lambda^{2/3}\phi_0}{\phi_0^2}\quad,\nonumber
\end{eqnarray}
where $\lambda= \lambda_1/\lambda_4$, making the approximate solution
\begin{eqnarray}
  \phi(t)&=&\left\{\begin{array}{cc}
    \phi_{L1}(t),&t<t_*\quad,\\
    \phi_{L2}(t),&t>t_*\quad.
    \end{array}
  \right.
\end{eqnarray}
Plugging in some numbers $\lambda_1=10^{-10}, \lambda_4=1,
\phi_0=-0.01$, we find that the analytical approximation fits quite
well to the numerical results, see Figure
\ref{fig:linear_plus_quartic:phase_space}. The terminal speed at time
$t_f$, i.e. at the bottom of the potential, is not clearly visible in
this plot, but evaluates to $\dot{\phi}(t_f)=1.22\times10^{-7}$
(analytic) which agrees quite well with the numerical result
$\dot{\phi}(t_f)=1.35\times10^{-7}$.

Finally, we can now estimate the amount of overshoot. For this purpose
we need to determine the time when the field reaches the bottom,
$t_f$, and its velocity there, $\dot\phi_{L2}(t_f)$, by evolving the
approximate solution Eq.~\eqref{eq:linear_plus_quartic:analytic} for
until $\phi=0$.

One then uses $\phi_R(t_f)=0,\;\dot\phi_R(t_f)=\dot\phi_{L2}(t_f)$ as
initial conditions for the solution to Eq.~\eqref{eq:eom:linear} on
the shallow inflationary slope on the right of the matching point
$\phi=0$. The overshoot, i.e. the field position $\phi_R(t_c)$ at time
$t_c=\sqrt{3/V_-}$ when curvature becomes subdominant to the vacuum
energy driving inflation, evaluates to (only neglecting terms
containing $V_-$)
\begin{equation}
  \phi(t_c)=\frac{3}{4\sqrt 2}\,\sqrt{\epsilon}-\left(\frac{\phi_*}{\phi_0}\right)^{3/2} \left(4 - 3 \frac{\phi_*}{\phi_0}\right)^{1/2} \phi_0\quad.
\end{equation}
Writing it in this form, we take the matching point $\phi_*$ as
independent parameter characterizing the two coefficients
$\lambda_1,\lambda_4$. If we move the matching point all the way to
the left, i.e. to $\phi_*=\phi_0$, we are dealing with a purely linear
potential and recover
\begin{equation}
  \phi(t_c)\stackrel{\phi_*\to\phi_0}{\to}\frac{3}{4\sqrt 2}\,\sqrt{\epsilon}-\phi_0\quad,
\end{equation}
in agreement with Eq.~\eqref{linear_overshoot}. If on the other hand we move
$\phi_*$ all the way to the right, we obtain
\begin{equation}
  \phi(t_c)\stackrel{\phi_*\to0}{\to}\frac{3}{4\sqrt 2}\,\sqrt{\epsilon}\quad,
\end{equation}
which is the formal limit of the overshoot for a quartic potential:
the field reaches the bottom of the potential at zero
speed. Neglecting terms with $V_-$ and setting $\dot{\phi}_f=0$, this
agrees with Eq.~\eqref{over_shoot}. Note that the overshoot is smaller
than the overshoot appearing in a purely linear potential. Indeed, it
is rather intuitive that the overshoot of the field in a binomial
potential can be at most as large as the overshoot derived from the
lowest-power monomial appearing in the scalar potential alone.

Finally, a trinomial can be approximate as the behavior of the
highest power monomial matched to the solution of the lower-power
binomial, and this procedure may be iterated to higher powers
still. Thus, we expect that the above argument concerning the upper
limit on the overshoot for the binomial case generalizes to the case
of full power-series expansion of the scalar potential in positive
powers of $\phi$.

\subsection{Binomial $V=V_{-}+\frac{1}{2}m^2\phi^2+\frac{1}{4}\lambda_4\phi^4$}
\begin{figure*}
  (a)\includegraphics[width=0.45\textwidth]{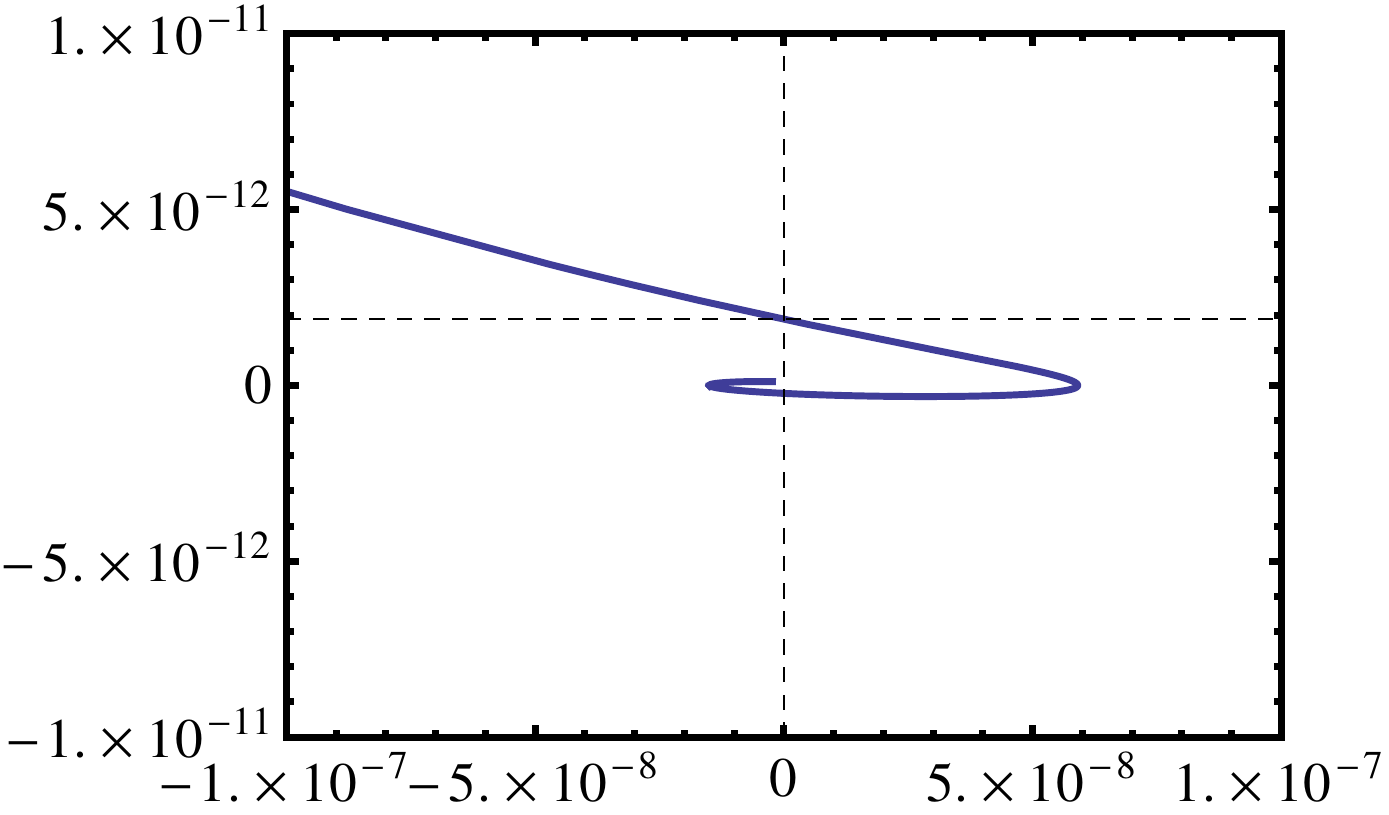}
  (b)\includegraphics[width=0.45\textwidth]{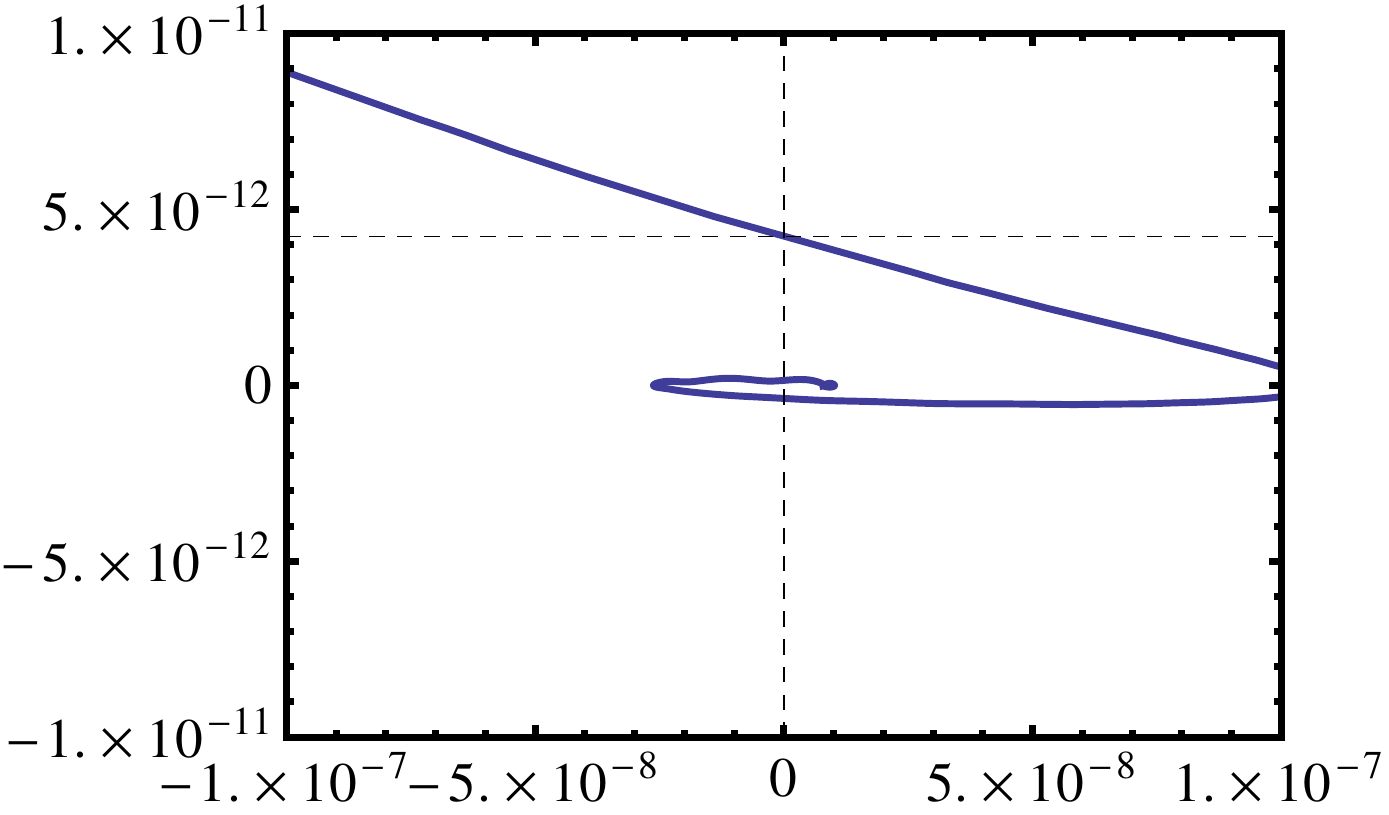}
  \caption{Phase space for a field rolling in the potential
    $V=\frac{1}{2}m^2\phi^2+\frac{1}{4}\lambda_4\phi^4$ (a) using the
    analytical expression Eq.~\eqref{eq:quadratic_plus_quartic:analytic}
    and (b) using numerical integration of the differential
    equation. The analytical approximation underestimates the terminal
    velocity by an order of magnitude (note the different scales on
    the $\dot{\phi}$ axes).}
  \label{fig:quadratic_plus_quartic:phase_space}
\end{figure*}

We follow the same strategy as in the previous section, matching
solutions where either quadratic part or the quartic part dominate to
obtain
\begin{eqnarray}\label{eq:quadratic_plus_quartic:analytic}
  \phi_{L1}&=&\frac{8\phi_0}{8+ \lambda_4\phi_0^2 t^2}\quad, \nonumber\\
  \phi_{L2}&=&\frac{4\pi\alpha(1-\alpha)}{\sqrt{\lambda_4}t} Y_1(\alpha\sqrt{\lambda_4}\phi_0 t)\\
  &&\phantom{+}\times\left[_0F_1(1;2\alpha(\alpha-1))-\alpha~_0F_1(2;2\alpha(\alpha-1))\right]\nonumber\\
  &&+\pi\alpha^2\phi_0~_0F_1(2;-\frac{\alpha^2\lambda_4\phi_0^2}{4}t^2)\nonumber\\
  &&\phantom{+}\times\left[2(1-\alpha)Y_0(-2\sqrt{2\alpha(1-\alpha)})\right.\nonumber\\
    &&\phantom{+\times}\left.+\sqrt{2\alpha (1-\alpha)}Y_1(-2\sqrt{2\alpha(1-\alpha)})\right]\quad,\nonumber
\end{eqnarray}
where $\alpha=-\frac{m}{\sqrt{\lambda_4}\phi_0}, 0<\alpha<1$. Figure
\ref{fig:quadratic_plus_quartic:phase_space} shows the phase space of
this solution, with panel (a) using the analytical expression of Eq.~\eqref{eq:quadratic_plus_quartic:analytic} and (b) using numerical
integration of the differential equation
$\ddot{\phi}+\frac{3}{t}\dot{\phi}+m^2\phi+\lambda_4\phi^3=0$. We chose
$\lambda_4=0.1, \phi_0=-0.1, m=10^{-5}$. 
The analytical estimate of terminal velocity is within the right order
of magnitude of the numerical result.

In order to estimate the amount of overshoot like in the last
subsection, we need to find solve $\phi_{L2}(t)=0$ with $\phi_{L2}$
from Eq.~\eqref{eq:quadratic_plus_quartic:analytic}. This can only be
done numerically. Thus we unfortunately cannot give an analytic
estimate for the overshoot.

We note that the better agreement of analytics and numerics in the
case of previous polynomial of linear and quartic is caused by the
fact that the transition region between the quartic and linear part is
much shorter than the one between the quartic and quadratic part. For
$|\phi|<1$, the range in $\phi$ where $\phi^3\lambda_4
\approx\lambda_1$ is much smaller than for $\phi^3\lambda_4 \approx
m^2\phi$.  

\section{Non-zero initial speed -- generalization to the landscape}\label{sec:nonzero_initial_speed}

The discussion so far established parametrically small or even
vanishing overshoot in a monotonically falling non-inflationary but
otherwise arbitrary potential after tunneling leading into a slow-roll
inflationary plateau using the single-field CDL boundary condition of
zero initial speed $\dot\phi_{0}=0$. Here we will give a simple
argument which generalizes these results to arbitrary finite initial
speed $\dot\phi_{0}\ne 0$ as long as the post-tunneling evolution
starts in a bubble geometry dominated by negative spatial curvature.

Assume the scalar field to emerge from CDL tunneling but with finite
$\dot\phi_{0}\ne 0$. For sufficiently early times $\epsilon \ll 1$ the
friction term in the curvature-dominated equation of motion will
completely vanquish the force term $\partial V/\partial\phi$. This
leaves at very early times an equation of motion
\begin{equation}
  \ddot{\phi}=-\,\frac{3}{t }\dot{\phi}\quad,
\end{equation}
which is solved by
\begin{equation}
  \dot{\phi}(t)=\dot{\phi}(\epsilon)\,\left(\frac{\epsilon}{t}\right)^{3}\quad.
\end{equation}
This implies that a given finite initial speed
$\dot{\phi}_{0}=\lim\limits_{\epsilon\to 0} \dot{\phi}(\epsilon)$ will
fall to a given fraction $\alpha\equiv\dot{\phi}(t)/\dot{\phi}_{0}$ in
an arbitrarily short time
\begin{equation}
  t_{\alpha}=\frac{\epsilon}{\alpha^{1/3}}\quad,
\end{equation}
once we take $\epsilon\to 0$.

Thus a given finite initial speed will approach a value arbitrarily
close to zero in arbitrarily short time. The solutions to monomial
potentials with finite initial speed asymptote to the our results with
zero initial speed at arbitrarily early times. Thus, in a bubble with
negative spatial curvature, the overshoot results hold for initial
conditions of arbitrary finite initial velocity $\dot{\phi}_{0}$ as
well.

There are immediate consequences of this universal behavior for the
multi-field situation in a landscape of local minima. The path of a
CDL bounce with minimum action will generically not be one that leads
into a basin of classical attraction (determined by the gradient of
the scalar potential $\vec{\nabla}\phi$ at the point of exit from
tunneling) towards a slow-roll inflationary region. 

For instance, the minimum action bounce may very well lead us directly
into our present day vacuum, bypassing the inflationary phase
completely. Conversely, a CDL bounce may exit into a basin of
classical attraction towards a slow-roll inflationary region with a
higher-action bounce. However, a bubble universe created this way will
undergo rapid inflation. We will not enter the discussion of how to
weigh different CDL bounces leading to different amounts of slow-roll
inflation post-tunneling. We merely require the existence of $\sim 60$
e-folds of slow-roll as an anthropic post-selection criterion to
select against otherwise smaller bounces leading directly into the
final vacuum.

This instanton is generically a curved trajectory in scalar field
space which provides some of the scalar fields with a finite initial
speed after tunneling. As the initial speed vector may very well point
along the inflationary direction in scalar field space, this would
generically lead to overshoot of the inflationary plateau region if
small or vanishing overshoot were contingent on vanishing initial
speed.

Thus the overshoot problem is not as severe even for finite initial
velocities. This provides a crucial generalization to a multi-field
landscape: If the landscape is populated via CDL tunneling -- and this
is the only known and controlled mechanism so far -- then there never
is a large overshoot problem on the \emph{first} slow-roll
inflationary plateau reached via steepest-decent after the exit from a
CDL tunneling event. Thus, on reaching this first plateau after exit
from CDL tunneling, small-field and large-field inflation regions in
the landscape are on equal footing with respect to the phase space of
initial conditions.

\section{Conclusions}\label{sec:conclusions}
We have shown that the overshoot problem in inflation after tunneling,
i.e. inflation in open universes, is not as severe as it might first
seem. We have demonstrated this for arbitrary monomial exit potentials
$V(\phi)=\frac{\lambda_n}{n}\phi^n$ -- describing the steep downhill
part of the potential at $\phi<0$ after emerging from tunneling --
turning into an inflationary plateau with a linear potential for
$\phi=0$ with its slope given by the first slow-roll parameter
$\epsilon$. In this setup we have found that for the three
lowest-power monomial exit potentials $\phi,\phi^2,\phi^3$ there is a
finite amount of overshoot
\begin{equation} \label{overshoot_end}
  \Delta\phi_{CDL}=\frac{3}{4\sqrt 2}\sqrt\epsilon+{\cal O}(1)|\phi_0|\quad.
\end{equation}
Surprisingly, for monomials $\phi^n\;,\;n\geq 4$ we get no overshoot
at all as the field already enters the inflationary phase on the exit
potential. The larger $n$, i.e. the steeper the exit potential and the
faster it asymptotes to slow-roll flatness, the more time the field
spends on the flat part of the exit potential. In addition, the Hubble
friction $1/t^2$ will be larger as well. These two effects combined
result in a more and more efficient slow down until for $n\geq4$, the
field comes to a complete stop.

Next we generalized our analysis to include polynomial exit
potentials. For a binomial, the overshoot is controlled by $|\phi_0|$
and determined by the lowest power monomial. This property carries
directly over to a general exit potential written as a power series
expansion around the point $\phi=0$ where slow-roll flatness has set
in.

Finally, we have shown that this absence of a parametrically large
overshoot for small-field inflation is actually \emph{independent} on
the initial speed $\dot\phi_0$ of the field. This seems to be in
contradiction with a comment made in~\cite{Bird:2009pq} that adding
negative curvature in general can be easily overwhelmed by large
enough initial field speed. However, our results are actually
compatible with this comment: the specific boundary condition of the
CDL instanton, namely that $a(t)\to t$ for small t, is crucial for
having curvature domination early on. This boundary condition was not
used in the comment of~\cite{Bird:2009pq}.

In the limit of small inflationary energy scale compared to the
tunneling scale, the CDL tunneling boundary conditions for the
negative spatial curvature and the scale factor are the only
conditions for our results to be valid. However, this implies that our
estimates of the overshoot generalize immediately to the case of CDL
tunneling in a multi-field landscape. There, some of the scalar fields
will generically have finite initial speed after tunneling. 

In the context of the string landscape populated by CDL tunneling, our
analysis shows that small-field and large-field inflation have
parametrically the same volume of phase space of initial conditions.

\section*{Acknowledgements} 
AW is grateful to L.~McAllister and S.~Raby for several illuminating
comments and discussions. The authors wish to thank N.~Kaloper for
valuable comments on the draft. This work was supported by the Impuls
und Vernetzungsfond of the Helmholtz Association of German Research
Centers under grant HZ-NG-603, and German Science Foundation (DFG)
within the Collaborative Research Center 676 ``Particles, Strings and
the Early Universe''.

\bibliographystyle{kp}
\bibliography{the_overshoot_problem}
\end{document}